\documentclass[10pt,preprint,a4paper]{aastex}

\usepackage{graphics,epsf}
\usepackage{amsmath}                
\usepackage{amsfonts}               
\usepackage{amssymb}                
\usepackage{epsfig}                 

\def \kev{\rm{keV}}

\def \cm{~\rm{cm}}
\def \s{~\rm{s}}
\def \km{~\rm{km}}

\def \g{~\rm{g}}

\def \AU{~\rm{AU}}
\def \erg{~\rm{erg}}

\def \yr{~\rm{yr}}
\def \pc{~\rm{pc}}

\def \days{~\rm{day}}

\title{MERGERBURST TRANSIENTS OF BROWN DWARFS WITH EXOPLANETS}

\author{Ealeal Bear\altaffilmark{1} Amit Kashi\altaffilmark{1} and Noam Soker\altaffilmark{1}}

\altaffiltext{1}{Department of Physics, Technion -- Israel Institute of
Technology, Haifa 32000 Israel; ealealbh@gmail.com; kashia@physics.technion.ac.il;
soker@physics.technion.ac.il.}

\begin{document}

\begin{abstract}
We explore the properties of an optical transient event formed by the
destruction of a planet by a brown dwarf (BD) -- a BD-planet mergerburst.
When a massive planet approaches a BD towards a merging process it will be tidally
destroyed and will form an accretion disk around the BD.
The viscosity in the disk sets the characteristic time for the event -- several days.
We suggest that BD-planet mergerburst events have light curves resembling those of other
intermediate luminosity optical transient (ILOT) events, such as V838 Mon, but at shorter timescales and lower luminosities.
With the high percentage coverage of the sky, we expect that such events will be detected in the near future.
\end{abstract}



\section{INTRODUCTION}
\label{sec:intro}
Brown dwarfs (BD) are substellar objects predicted by Kumar (1963; he termed them black dwarfs) about 50 years ago,
within the mass range of $\sim 13~{\rm M_J}$--$80~{\rm M_J}$ (e.g., Johnson et al. 2011).
BDs never burn Hydrogen to Helium like main sequence (MS) stars
do, and generally experience gravitational contraction, which reduces
their radius as their age increases (e.g., Jorgens 2005), eventually reaching average densities larger
than those of solar type main sequence (MS) stars.
BDs are known to exist also in binary systems, both as the less massive component and
as the primary (e.g., Thies et al. 2010; Kraus et al. 2011).
In our study we focus on BDs hosting a planetary system.

The BDs formation mechanism is still under debate (Luhman et al.
2007), with a few possible channels. In the first channel the BD
is formed via a stellar like mechanism, where gravitational
collapse is followed by accretion from a circumstellar disk (Thies
et al. 2010; Monin et al. 2010). In this channel we expect that many BDs will host planetary systems.
The second channel is commonly
referred to as the `ejection channel', in which BDs are ejected
from the disk, and hence are not fully developed. Therefore BDs
are similar to stellar-embryos (Reipurth \& Clarke 2001). Other
channels suggest less commonly used explanations such as disk
fragmentation (for more details see Whitworth et al. 2006).
Although their formation process is still under debate, and despite the
so called brown dwarf desert, some claim that it may be possible that
BDs are much more common than MS stars (e.g., Reid 1999).

Furthermore, BDs are known to exist in star forming regions and in
nearby galactic fields (e.g., Joergens 2008, 2010; Caballero 2008;
Luhman 2006; Bonnell et al. 2003). A $5~{\rm M_J}$ planet
orbiting a $25~{\rm M_J}$ young BD at an orbital separation of
$55\AU$ has already been discovered, at a distance of $70 \pc$
(Chauvin et al. 2004; Allers et al. 2010 and references
therein). Over all, the evolution of planetary systems around
BD deserve attention. We study the observational signature of a
planet merging with a BD. Even in globular clusters such a
processes might occur. Although planets are rare in globular
clusters and hard to detect (Bonnel et al. 2003), the high stellar
densities might lead to BD-planet merger if both BDs and planets
exist in globular clusters.

A key parameter in our proposed scenario is that BDs have typical
radii like those of Jupiter-like planets (e.g., Kumar 1963;
Chabrier \& Baraffe 2000; Mohanty et al. 2004; Cruz
2007), and hence they are much denser than massive planets.
This implies that in a close encounter the planet will be tidally
destructed by the BD.

In this paper we propose a scenario of a planet and a BD merger
that creates a burst -- a BD-planet mergerburst.
The planet can be a free floating planet in dense globular or open clusters (Soker et al. 2001),
or a close planet that was born with the BD.
This mergerburst event is likely to be observed as a several days burst in the visible and IR bands.
A shorter peak in the UV and X-ray bands might be presence as well.
The energy source is accretion of matter to the BD from an accretion disk formed from the
destructed planet.
Merger events between planets have been discussed before (e.g., Zhang \& Sigurdsson 2003) and are expected to
be super Eddington events that release extreme-UV and possibly X-ray radiation for a few hours.
Our setting is different than what was studied by Zhang \& Sigurdsson (2003), as we consider tidal destruction and the
formation of an accretion disk.

In Section \ref{sec:Tidal} we discuss tidal interactions and
outline the mergerburst process. In Section \ref{sec:destructionprocess} we review
the physical properties for this process, and in section \ref{sec:observations} we discuss its observational
properties.
We summarize in section \ref{sec:Summary}.

\section{TIDAL DESTRUCTION OF A PLANET BY A BROWN DWARF}
\label{sec:Tidal}

Previous studies of tidal interaction of planets with MS stars that
lead to merger events have been conducted, e.g., Retter et al. (2007)
who discussed a planet capture model for V838~Mon.
However, this model suffers from some fundamental problems (Soker \& Tylenda 2007). The
main one is that most of the orbital energy is released deep in the envelope of the MS star.
In such a case, most of the energy is channeled to uplift inner shells of the envelope,
rather that to radiation and mass ejection.
As shown by Soker \& Tylenda (2007) this process does not cause much increase in luminosity.

A mergerburst between a planet and a BD is an example of an extreme case of an eruptive
mass transfer event. The formation of an accretion disk becomes more efficient as
the density ratio between the accretor and the destructed companion increases.
In cases where the densities ratio is insufficient to
destruct the companion, a direct merger is likely to occur.
The mergerburst process is not the same as mass transfer within a binary system, or mass transfer from a
very large and young protostellar disk. The latter processes can not supply the required energy
in such a short time (Soker \& Tylenda 2003; Tylenda \& Soker 2006).

We follow the V838~Mon mergerburst model of Soker \& Tylenda
(2007; see also Soker \& Tylenda 2003, 2006; Tylena \& Soker
2006), where a MS secondary of $\sim 0.3~{\rm M_\odot}$ merged
with a MS primary of $\sim 8~{\rm M_\odot}$.
This scenario explains the peak in luminosity which was observable for
$~100\days$ in the visible band.
We here suggest that the BD-planet mergerburst
is similar in many aspects to V838, e.g., in light curve.
A similar accretion process onto
a MS has been suggested before (e.g., Fujimoto \& Iben 1989;
Shaviv \& Starrfield 1988), but in different settings.

In our model we downscale the masses of the two stars that
participated in the eruption of V838~Mon, according to the
accepted model, by a factor of $\sim 100$.
to match the BD-planet scale.
We take a BD with a typical mass of $M_{\rm
BD}=60~{\rm M_J}$, and a planet of $M_{\rm p}=3~{\rm M_J}$. These
masses will be the typical masses used for these objects
throughout this paper, with allowed primary mass range of $\sim
10$--$120~{\rm M_J}$ and secondary mass range of $\sim 1$--$13~{\rm M_J}$.

We require our system to be old for the BD to be compact.
As BDs do dot expand as they age, to bring the planet closer to the BD the orbit must be perturbed by an external source
(a passing star or a third body in the system).
For that, the expected orbit when tidal interaction takes place near periastron passage is highly eccentric, and there is no synchronization.
Tidal destruction rather than Roche lobe over flow is the relevant process.

Our scenario is based on BDs being considerably denser than MS stars.
Whether or not a planet is destructed by tidal forces when
approaching another object, here it is a BD, is given by the condition that the tidal shredding
radius of the planet $R_s$ (e.g., Nordhaus et al. 2010) be larger than the radius of the BD
\begin{equation}
R_{\rm s}\simeq R_{\rm p}\left(\frac{2 M_{\rm BD}}{M_{\rm p}}\right)^{1/3}
\simeq 3.4 \left(\frac{R_{\rm p}}{1~{\rm R_J}}\right)
\left(\frac{M_{\rm p}}{3~{\rm M_J}}\right)^{-1/3}
\left(\frac{M_{\rm BD}}{60~{\rm M_J}}\right)^{1/3} {\rm R_J} > R_{\rm BD}.  
\label{R tidal shred}
\end{equation}
The radius of a not-too young BD satisfies $R_{\rm BD} < R_s$, and the planet will be tidally destructed.
As the destructed planet is expected to have a considerable amount of angular momentum, it is
likely to form a disk around the BD.
For comparison, if we consider a solar mass MS star with a typical density of $\rho_{\rm MS} \sim 1.4\g \cm^{-3}$,
no substantial disk will be formed, since the shredding radius is too close to, or even smaller than,
the stellar radius.

\section{PHYSICAL PROPERTIES OF THE DESTRUCTION PROCESS}
\label{sec:destructionprocess}

We now turn to estimate the available energy for the BD-planet
mergerburst. This will serve us later in estimating the
observational properties of such an event.
The energy released is
\begin{equation}
E_{\rm acc} \simeq \frac{1}{2}\frac{GM_{\rm BD}M_{\rm p}}{R_{\rm BD}}
=3 \times 10^{45} \left(\frac{M_{\rm BD}}{60~{\rm M_J}}\right)
\left(\frac{M_{\rm p}}{3~{\rm M_J}}\right)
\left(\frac{R_{\rm BD}}{1~{\rm R_J}}\right)^{-1} \erg. 
\label{eq:Eacc}
\end{equation}
This value is $\sim 300$ times less than Kashi \& Soker (2011) get for
calculating the eruption energy of V838~Mon [$E_{\rm acc}(\rm V838~Mon) \simeq 9 \times 10^{47} \erg$; see equation \ref{eq:EaccV838Mon} below].
We note that Tylenda \& Soker (2006) estimated the total energy of V838~Mon to be $\sim 0.3$--$10 \times 10^{47} \erg$,
taking into account the energy expected to be in a polytropic envelope together with the kinetic and radiated energy.

The accretion time must be longer than the viscosity time scale
for the accreted mass to lose its angular momentum.
The viscous timescale is (e.g., Dubus et al. 2001)
\begin{equation}
\begin{split}
t_{\rm{visc}} &\simeq \frac{R_{\rm BD}^2}{\nu}
\simeq 2.5
\left(\frac{\alpha}{0.1}\right)^{-1}
\left(\frac{H/R_{\rm BD}}{0.1}\right)^{-1}
\left(\frac{C_s/v_\phi}{0.1}\right)^{-1}
\left(\frac{R_{\rm BD}}{1~{\rm R_J}}\right)^{3/2}
\left(\frac{M_{\rm BD}}{60~{\rm M_J}}\right)^{-1/2} \days, 
\end{split}
\label{eq:tvisc1}
\end{equation}
where $\nu=\alpha C_s H$ is the viscosity of the disk parameterized with the disk$-\alpha$ parameter,
$H$ is the vertical thickness of the disk, $C_s$ is the sound speed and $v_\phi$ is the Keplerian velocity.
We scale $M_{\rm BD}$ and $R_{\rm BD}$ in equation (\ref{eq:tvisc1}) according to the parameters of a $60~{\rm M_J}$ BD.
Shortly after the destruction of the planet, the gas is hot and has no time to settle into a thin disk;
a thick disk is likely to be formed around the BD.
For that, the value of the disk$-\alpha$ parameter is highly uncertain.

The timescale for the eruption is expected to be $\sim {\rm few} \times t_{\rm{visc}}$.
The masses of the BDs and planets that produce the transient event are expected to be
$M_{\rm BD} \simeq 13$--$80~{\rm M_J}$ and $M_{\rm p} \simeq 1$--$10~{\rm M_J}$, respectively.
The proposed scenario for BDs works for very low mass MS stars as well.
However, the star cannot be much more massive than $\sim 0.1 {\rm M_\odot}$, as its density decreases
to the point that direct merging occurs before tidal destruction takes place.

\begin{figure*}
\resizebox{1.0\textwidth}{!}{\includegraphics{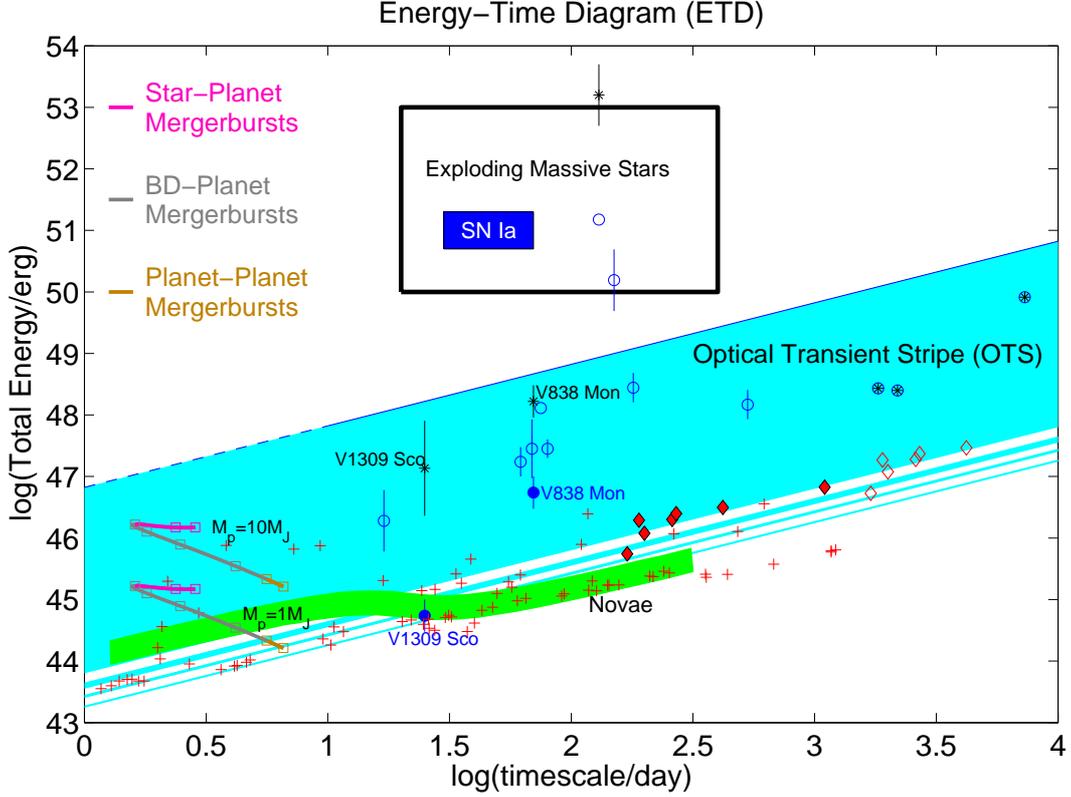}}
\caption{\footnotesize{ The calculated location of the BD-planet
mergerbursts as well as observed transient events on the
Energy-Time Diagram (ETD; taken from Kashi \& Soker 2011). Blue
empty circles represent the total (radiated plus kinetic) energy
of the observed transients $E_{\rm{tot}}$ as a function of the
duration $t$ of their eruptions (generally defined as the time it
took the transient to decrease by 3 magnitudes in the V-band).
The Optical Transient Stripe (OTS), is a constant luminosity region in the ETD.
It is populated by accretion powered events such as ILOTs (including mergerbursts),
major LBV eruptions, and the new group of BD-Planet mergerbursts we predict in this paper.
Blue filled circles represent observed transients that are
mergerbursts. The green line represents nova models
computed using luminosity and duration from Kulkarni et al.
(2007). Nova models from Yaron et al. (2005) are marked with red
crosses, and models from Shara et al. (2011) are represented with diamonds.
The total energy does not include the energy which is
deposited in lifting the envelope that does not escape from the
star. Where a model exists to calculate the total gravitational
energy released by the accreted mass, the available energy, it is
marked by a black asterisk above or overlapping the blue circle.
For further details and the designation of the individual
transients see figure 1 of Kashi \& Soker (2011). The BD-Planet
mergerbursts are marked with two lines
(in the lower-left side of the figure), for two limiting cases of
a $1~{\rm M_J}$ (bottom line) and a $10~{\rm M_J}$ (upper line)
planets. The squares on each line are for accreting primaries with
mass of $10,~13,~20,~40,~60,~80,~100,$ and $120~{\rm M_J}$. The
colors of the lines represent primaries which are massive planets
themselves ($10$--$13~{\rm M_J}$), BDs ($13$--$80~{\rm M_J}$), and
very low mass MS stars ($80$--$120~{\rm M_J}$), according to the
legend. }} \label{fig:BDP_ETD}
\end{figure*}

Let us place the proposed scenario in the general picture of
merging stars and other binary systems with eruptive mass transfer
events. The physical process in both cases is similar, only
that a mergerburst is considered to be an extreme case of a mass
transfer event. The Energy-Time Diagram (ETD; Kashi \& Soker
2011) is a tool to help understand transient events. In the ETD,
presented here in Fig. \ref{fig:BDP_ETD}, the total energy of
transients is plotted as a function of their characteristic
timescale. The top of the ETD is the domain of the energetic
supernovae, and its bottom is populated by novae. The relevant
part of the ETD is the Optical Transient Stripe (OTS), populated
by Luminous Blue Variables (LBV) major eruptions, such as the
eruptions of $\eta$ Car and P Cyg, and by Intermediate Luminosity
Optical Transients (ILOTs; also called Intermediate Luminosity Red Transients),
such as V838~Mon, M85-2006OT, M31-RV
and V1309~Sco. Intermediate Luminosity Optical Transients
are a new type of objects observed in the last two decades.
Observationally, they are defined as events with luminosities
between nova eruptions and supernova (SN) explosions, in an
absolute magnitude range $-17<M_V<-8$, strongest in the
visible or IR bands, and decaying as red objects (e.g., Berger et
al. 2009; Kulkarni et al. 2007). The OTS has a constant slope
of 1 in the $\log(E_{\rm{tot}})$--$\log(t)$ plane, and extends
over $\sim3$ orders of magnitude. This implies a constant average
luminosity (over the duration of the eruptions) of
$L_{\rm{stripe}} \simeq 4 \times 10^{40 \pm 1}~\erg \s^{-1}$.
Kashi \& Soker (2011) suggested that being on the same stripe, and
showing other similar properties, hint that the eruptions on the
OTS have the same energy source -- an eruptive accretion event,
possibly through an accretion disk.

We present the calculated BD-planet Mergerburst on the ETD of Fig.
\ref{fig:BDP_ETD}. The BD-Planet mergerbursts are marked with two
lines, for two limiting cases of a planet with a mass of $1~{\rm
M_J}$ (bottom line) and $10~{\rm M_J}$ (upper line). The colors of
the lines represents primaries which are massive planets
themselves ($10$--$13~{\rm M_J}$), BD primaries ($13$--$80~{\rm
M_J}$) and low mass stars ($80$--$120~{\rm M_J}$). The squares on
each line are for primaries mass of
$10,~13,~20,~40,~60,~80,~100,~120~{\rm M_J}$. The shape of the
line is determined from the varying profile of $R_{\rm BD}(M_{\rm
BD})$, taken from figure 1 in Burrows et al. (2011). The 5 Gyr BD
profile we adopted from figure 1 in Burrows et al. (2011) shows
decreasing BD radius up to a minimum radius of $0.78~{\rm R_J}$ at
$74~{\rm M_J}$, follows by an increase for more massive BDs. This
means that the timescale, taken from equation (\ref{eq:tvisc1}),
decreases for BD masses up
to $\sim 80~{\rm M_J}$ and then increases. 
The energy also increases up to $\sim 80~{\rm M_J}$, and then slightly decreases because the
radius of the BD increases faster than its mass in that range.

We find that the BD-Planet mergerbursts are located on the left
continuation (lower timescales and energies) of the OTS. Hence, we
suggest that the expected observed outburst of BD-Planet
mergerbursts are scale-down version of ILOTs such as V838~Mon and
V1309~Sco (Mason et al. 2010; Tylenda et al. 2011a; Stepien 2011). The BD-planet
mergerbursts form the lower part of the OTS.

\section{OBSERVATIONAL CONSIDERATIONS}
\label{sec:observations}

\subsection{Visible light}
\label{sec:observations_visible}

As to our best knowledge BD-Planet mergerbursts have never been observed, it is in place to
predict the expected observational properties of a typical BD-Planet Mergerburst.
We start by suggesting the typical optical light curve.

Kashi \& Soker (2011) and Kashi et al. (2010) showed that the V-band light curve of
ILOTs and LBVs from maximum to $\sim 4$ magnitude below maximum are similar up to time-rescaling.
There is yet no accurate model that explains the particular
shape, but nevertheless its similarity is clearly not a coincidence.
The time scale for the decline depends on the accreting object,
and on the properties of the accretion disk.

\begin{figure}[t]
\resizebox{0.7\textwidth}{!}{\includegraphics{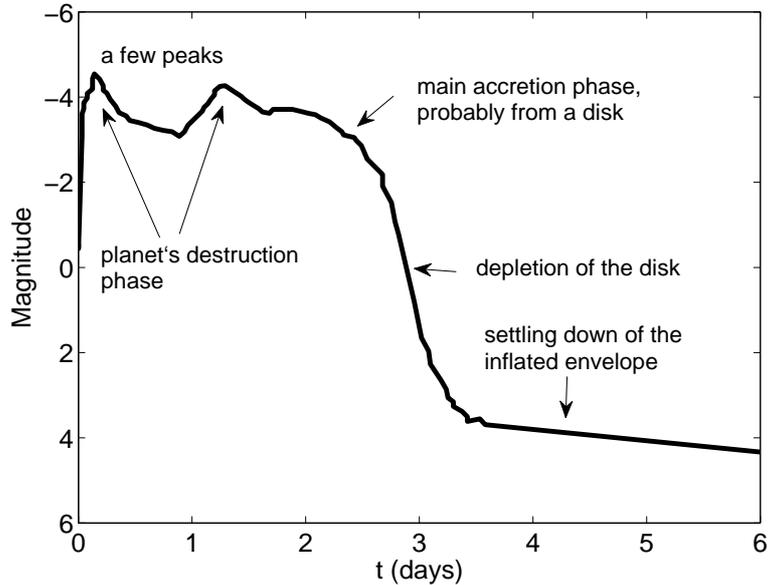}}
\caption{\footnotesize{
A schematic drawing of the crudely expected V-band light curve of a BD-plant mergerburst.
The light curve is obtained by calibrating the light curve of V838~Mon to a less luminous BD-Planet Mergerburst,
taking energy according to equations (\ref{eq:Eacc}) and (\ref{eq:Lmax}), and timescale from equation (\ref{eq:tvisc1}).
The light curve is presented in absolute magnitude scale, and is scaleable for different BD and planet masses according to
equations (\ref{eq:Lmax}) and (\ref{eq:tvisc1}).
As is observed in other ILOTs, it is expected that at the beginning of BD-Planet mergerbursts there will be one or more peaks,
followed by a decline as plotted.}}
\label{fig:BDPlightcurve}
\end{figure}

We therefore suggest that the light curve of BD-Planet
mergerbursts resemble the light curve of ILOTs. We take
V838~Mon as a typical ILOT light curve and scale it down to a
BD-Planet mergerbursts light curve. The typical timescale for the
decline of the light curve of V838~Mon was $\tau(\rm V838~Mon) \simeq 70 \days$ (Sparks et al. 2008).
The total available energy in the eruption of V838~Mon was (Kashi \& Soker 2011)
\begin{equation}
\begin{split}
E_{\rm acc}(\rm V838~Mon) &= \frac{1}{2}\frac{G M_1 M_2}{R_1}
\simeq 9 \times 10^{47} \left(\frac{M_1}{8 ~{\rm M_\odot}}\right)
\left(\frac{M_2}{0.3 ~{\rm M_\odot}}\right)
\left(\frac{R_1}{5 ~{\rm R_\odot}}\right)^{-1} \erg. 
\end{split}
\label{eq:EaccV838Mon}
\end{equation}
This energy is much larger than the radiated energy of $\simeq 2.5 \times 10^{46} \erg$ (Tylenda 2005),
as it includes also the kinetic energy of the ejecta and the energy that went to inflate the envelope.
Substituting typical values of V838~Mon (equation \ref{eq:EaccV838Mon}),
the total available energy ratio between a BD-Planet Mergerburst and V838~Mon, derives from equations (\ref{eq:Eacc}) and (\ref{eq:EaccV838Mon}) is
\begin{equation}
\begin{split}
\chi = \frac{E_{\rm acc}(\rm{BD-planet})}{E_{\rm acc}(\rm V838~Mon)}
\simeq \frac{1}{300} \left(\frac{M_{\rm BD}}{60~{\rm M_J}}\right)
\left(\frac{M_{\rm p}}{3~{\rm M_J}}\right)
\left(\frac{R_{\rm BD}}{1~{\rm R_J}}\right)^{-1}. 
\end{split}
\label{eq:chi}
\end{equation}
{From} the lack of an accurate model, we take the ratio between the total radiated energy in the V-band of a BD-Planet Mergerburst
and V838~Mon to be $\sim \chi$ as well.
Under this assumption we calibrate the V-band light curve of a typical BD-Planet Mergerburst (Fig. \ref{fig:BDPlightcurve}),
by taking the light curve of V838~Mon, $L_{\rm V838~Mon}(t)$, multiply it by the energy ratio $\chi$ and divide it by the timescale ratio
\begin{equation}
L(\rm BD-planet)(t) = \chi L_{\rm V838~Mon}(t) \frac{\tau(\rm V838~Mon)}{t_{\rm{visc}}(BD-planet)}.
\label{eq:Lbdp}
\end{equation}
Using the viscous timescale from equation \ref{eq:tvisc1} we find that the expected peak luminosity is
\begin{equation}
\begin{split}
L(\rm BD-planet)_{\rm max} \approx
& 10^{37-38} \left[\frac{\tau(\rm V838~Mon)}{70 \days}\right]
\left(\frac{\alpha}{0.1}\right)
\left(\frac{H/R_{\rm BD}}{0.1}\right)
\left(\frac{C_s/v_\phi}{0.1}\right)\\
& \times \left(\frac{M_{\rm BD}}{60~{\rm M_J}}\right)^{3/2}
\left(\frac{M_{\rm p}}{3~{\rm M_J}}\right)
\left(\frac{R_{\rm BD}}{1~{\rm R_J}}\right)^{-5/2} \erg \s^{-1}. 
\end{split}
\label{eq:Lmax}
\end{equation}
As noted by Tylenda \& Soker (2003), the merger process is
complicated and can have different epochs with changes in luminosity.
Therefore, the predicted light curve plotted in Fig. \ref{fig:BDPlightcurve} is a schematic one.
We note here that a super Eddington luminosity from an accretion disk can be accounted for theoretically
(Dotan \& Shaviv 2010).

\subsection{UV and X-ray}
\label{sec:observations_xray}

A BD-planet mergerburst event can be an X-ray source during, as well as a long time after the event.
The Keplerian velocity around the BD is $\sim 350 (M_{\rm BD}/60~\rm{M_J})^{1/2} \km \s^{-1}$.
The inner boundary of the disk (in particular the boundary layer) might be a
source of UV emission, and accretion shock can lead to X-ray emission at $\sim 0.1 \kev$.
As well, an accretion disk might have an active corona.
However, the destruction of the planet is likely to eject gas in all directions.
This gas obscures the accretion disk that is expected to form during the main phase of the event.
On the other hand, the accretion disk might launch two jets,
that when shocked form hot gas that becomes an X-ray source.
Even if the X-ray source is as weak as
$L_{x-{\rm jet}} \sim 10^{-7} L_{\rm max} \sim 10^{30} \erg \s^{-1}$
it is not negligible. As it expected to last for hours to days, it might be detected.

Later, it is possible that some of the ejected gas falls back onto the BD and maintains the accretion disk.
By that time the gas around the BD is long dispersed and the accretion disk can be observed.
The accretion disk might become a strong source of UV radiation from its boundary layer
(Kashi \& Soker 2011), as well as an X-ray source from a disk-corona.

Another possible strong X-ray source is the BD itself. The accreted planet material will substantially spin-up the BD.
The BD becomes a magnetically active object. Accreting BDs might be strong (and peculiar)
X-ray sources, as is the case with FU~Tau (Stelzer et al. 2010).

There are two differences of BD-planet mergerburst events from MS-MS mergers such as V838 Mon.
($i$) The BD is not massive enough to hold a large envelope as a MS star does.
For example, a decade after its mergerburst event, V838 Mon is still a red giant (Tylenda et al. 2011b).
($ii$) The planet contains little mass, such that the optical deep phase will not last long.
Over all, if a BD-planet mergerburst event is detected, it should become
a prime target for X-ray (and UV) observations immediately with the detection, and for years to follow.

\subsection{Frequency of BD-planet mergerburst}
\label{sec:Frequency}

We can only give a crude estimate on the expected frequency of BD-planet mergerbursts.
We do so by comparing to known ILOTs formed from stellar mergers.
From the two known transients V838~Mon and V1309~Sco, we estimate that such events occur in systems
with masses in the range $\Delta M_{\rm ILOT} \approx 1$--$8~\rm {M_\odot}$
(Tylenda \& Soker 2006; Tylenda et al 2011; Stepien 2011) once every $\sim 10 \yr$ in our Galaxy.

The IMF in fig. 2 of Parravano et al.(2011) shows that the number of BDs and very low mass MS stars
that are considered here $\Delta M_{\rm BD-planet} = 10$--$120~\rm{M_J}$,
is $\sim 0.1$ times the number of stars in the mass
range $\Delta M_{\rm ILOT}\approx 1$--$8~\rm {M_\odot}$ that are thought to be the progenitor
of stellar ILOTs. 

We cannot estimate the frequency of BD-planet systems as to the
best of our knowledge there is only one known BD with an orbiting
planet (Allers et al. 2011; Chauvin et al. 2004). However, the
large number of BDs, and the large number of planets around stars
hint that BD with planetary systems might be very common. Namely,
we suggest that BD-planet merger are not much more rare that ILOTs
of MS stars. As we require the BD to be denser than the planet,
hence old enough, we do not expect BD-planet mergerbursts to occur
in young star-forming regions. However, in systems older than
$\sim 10^8 \yr$, mainly open clusters, the BDs are old
enough for their radius (Burrows et al. 1997, 2011) to fulfill
equation (\ref{R tidal shred}). In addition, in open clusters the
relatively high stellar density can lead to frequent perturbations
by surrounding stars that can cause planets to migrate toward the
BD. A prime target for such events might be open clusters.

Current surveys can easily detect novae in the Galaxy,
hence they will be able to detect BD-planet mergerbursts which
we predict to be even brighter (see Fig. \ref{fig:BDP_ETD}).
A problem might arise because once a BD-planet mergerburst is
observed, it might be confused with a nova over a short
observation cadence. The luminosity, color, and light
curve over a short time might be similar to those of some
novae, as it is might occur with stellar ILOTs (see Tylenda
\& Soker 2006). Large survey that reject ``uninteresting
events'', might mistakenly classify a BD-planet mergerburst as a
nova. However, follow up observations several weeks after maximum
can show that the event is not a nova. The main difference is that
the color will turn red instead of blue (see Tylenda \& Soker
2006).

\section{SUMMARY}
\label{sec:Summary}

In this paper we introduced the notion of a mergerburst between a planet and a brown dwarf (BD).
In this process the planet is shredded into a disk, and the accretion lead to an outburst. The destruction of
a component in a binary system and transforming it to an accretion disk is an extreme
case of mass transfer processes in binary systems. The destruction of the planet before it hits the
BD occurs because its average density is lower than that of the BD. We also require that the planet
enters the tidal radius. This can be caused in a highly eccentric orbit that is perturbed.
We assume that once the planet is destructed, the remnant of the merger behaves similarly to other
ILOTs, such as V838 Mon, but on a shorter time scale and with less energy.
This process is a super Eddington process.
A schematic presentation of the expected light curve is given in Fig. \ref{fig:BDPlightcurve}, with large
uncertainties in the exact time scale and luminosity. In any case, we expect such events to
occupy the lower part of the Optical Transient Stripe (OTS) on the Energy-Time diagram (ETD), as presented
in Fig. \ref{fig:BDP_ETD}.

BDs are denser than planets. This implies that planets are likely to be
tidally destructed before colliding with the BD. Due to the large angular momentum an accretion disk is
like to be formed.
The viscous time in the accretion disc, equation (\ref{eq:tvisc1}), is the main factor, but not the sole one,
that determines the decline time of the luminosity.
As with other mergerbursts, an extended envelope is expected to be formed.
Because of the low masses of the BD and the planet we expect the inflated envelope to be dispersed
in a typical time scale of months to years. Later accretion of fall back gas might form
(or maintain) an accretion disk, and be observed in the UV and X-ray bands.
The spun-up BD might become magnetically active, and a source of hard X-ray emission
for a very long time after the event.

Such a process is less likely to occur with solar type MS stars, as their density is lower. A planet is more likely to
merge with the star before being tidally destructed. The planet will deposit most of its energy deep in the envelope
of the MS star, and the outburst will be weak.
The high density of BDs make the merger with planets an interesting event.
With larger and larger coverage of the sky by more and more surveys, we expect the detection of such events in the
near future.
A favorable location for BD-planet mergerbursts might be open clusters where the BDs
are dense enough to result in a planet destruction.
As BD-planet mergerbursts may be confused with novae, observations at late times after the event are important for
distinguishing between them.
If an event with properties as outlined here in the figures is detected, it should immediately become
a prime target for X-ray observations.

We thank an anonymous referee for helpful comments.
AK acknowledges a grant from the Irwin and Joan Jacobs fund at the Technion.
This research was supported by the Asher Fund for Space Research
at the Technion, and the Israel Science foundation.

\end{document}